\documentclass[10pt,prd,aps,twocolumn,superscriptaddress,showpacs,floatfix,amsmath,amssymb]{revtex4}

\usepackage{bm}

\usepackage{amsmath}
\usepackage{graphicx}
\usepackage{amssymb}
\usepackage{mathrsfs}
\usepackage{bbm}

\newcommand{\be}{\begin{eqnarray}}
\newcommand{\ee}{\end{eqnarray}}

\begin{document}

\hskip .0cm
\title{{\large \bf A Gauge Field Theory  of Chirally Folded Homopolymers  \\
with Applications to Folded Proteins}}

\author{Ulf H. Danielsson}
\email{Ulf.Danielsson@physics.uu.se}
\affiliation{Department of Physics and Astronomy, Uppsala University,
P.O. Box 803, S-75108, Uppsala, Sweden}
\author{Martin Lundgren}
\email{Martin.Lundgren@physics.uu.se}
\affiliation{Department of Physics and Astronomy, Uppsala University,
P.O. Box 803, S-75108, Uppsala, Sweden}
\author{Antti J. Niemi}
\email{Antti.Niemi@physics.uu.se}
\affiliation{Department of Physics and Astronomy, Uppsala University,
P.O. Box 803, S-75108, Uppsala, Sweden}
\affiliation{
Laboratoire de Mathematiques et Physique Theorique
CNRS UMR 6083, F\'ed\'eration Denis Poisson, Universit\'e de Tours,
Parc de Grandmont, F37200, Tours, France}
\affiliation{Chern Institute of Mathematics, Tianjin 300071, P.R. China}


\begin{abstract}
We combine the principle of gauge invariance with 
extrinsic string geometry to develop a lattice model that can be
employed to theoretically describe properties of chiral, unbranched homopolymers. 
We find that in its low temperature phase the model is in the same
universality class with proteins that are deposited in the Protein Data Bank,   
in the sense of the compactness index. 
We apply the model to analyze various statistical aspects of  folded 
proteins. Curiously we find that it can produce results that are a very good
good match to the data in the Protein Data Bank.
\end{abstract}



\maketitle

\section{introduction}

Effective field  theory models are often employed and sometimes even with great success, 
to address complicated problems when the exact theoretical 
principles are either unknown, or have a structure that is far too complex 
for analytic or numerical treatments. Familiar examples of
powerful and predictive effective field theory models include the 
Ginzburg-Landau approach to superconductivity \cite{landau} and 
the Skyrme model of atomic nuclei \cite{skyrme}. 

In polymer physics field theory
techniques became popular after de Gennes \cite{degennes1}, \cite{degennes2}
showed  that the self-avoiding random walk 
and the $N \to 0$ limit of the $O(N)$ symmetric $(\phi^2)^2$ scalar 
field theory are in the same universality class in the sense of the compactness index.
The ensuing field theory approach is very powerful in characterizing  critical 
properties of homopolymers. However, to our knowledge there are no effective 
field theory models that allow for a detailed description of the geometry of collapsed,
chiral homopolymers. The goal of the present article is to develop such a model,
in the case of unbranched single-strand homopolymers.

\section{Compactness index}
 
We start by recalling the compactness index $\nu$ that describes how the radius of 
gyration $R_g$ scales in the degree of polymerization $N$
\be
R_g \ = \ \frac{1}{N} \sqrt{ \frac{1}{2} \sum_{i,j} ( {\bf r}_i 
- {\bf r}_j )^2 }  \propto \ L N^{\nu}
\label{nu}
\ee   
The value of $\nu$ is a universal quantity, in the limit of large $N$ \cite{degennes2}.
Here ${\bf r}_i$ ($i=1,2,...,N$) are the locations of the
monomers and $L$ is a form factor that characterizes 
an effective distance between monomers, it is not a universal quantity.
At high temperatures we expect that $\nu$ quite universally 
approaches the Flory value \cite{degennes2} $\nu \sim 
3/5$ that corresponds to the universality class of 
self-avoiding random walk; Monte-Carlo estimates
refine this to $\nu \approx 0.588 \dots$ \cite{sarw}. 
On the other hand, for collapsed polymers 
we expect to find $\nu \sim 1/3$. Since $\nu$ coincides with the
inverse Hausdorff dimension of the polymer, this means that a collapsed polymer
is as compact as ordinary matter. Finally, between the self-avoiding random walk phase and
the collapsed phase we expect to  have the $\Theta$-point that describes the universality class 
of a fully flexible chain. At the $\Theta$-point we expect $\nu \sim 1/2$.

\section{model}

We have found that the scaling law $\nu \sim 1/3$ of collapsed polymers 
can be  computed by the low temperature free energy of
a discrete version of the two dimensional Abelian Higgs model with
an $O(2) \sim U(1)$ symmetric Higgs field. In its three space dimensional version this
model was originally
introduced to describe superconductivity \cite{landau} and it has also found applications
in high energy physics, for example in the description of cosmic strings.
We first motivate this model in the present context by considering the continuum limit,  
where the polymer is approximated by 
a continuous one-dimensional string. The string is described by the position 
vector ${\bf r}(s)$ where the parameter $s \in [0,L]$ measures the 
distance along the string that has a total length $L$. The unit tangent 
vector of the string is
\[
{\bf t} = \frac{ d {\bf r}}{ds} .
\]
Together with the unit normal vector $\bf n$ and the unit binormal 
vector ${\bf b} = {\bf t} \times {\bf n}$ we have an orthonormal Frenet 
frame at each point along the string. In terms of the complex 
combination 
\[
{\bf e}^\pm_F =   {\bf n} \pm i {\bf b} 
\]
these three vectors are subject to the Frenet equations \cite{spivak}
\[
\frac{d{\bf t}}{ds} = \frac{1}{2} \kappa ({\bf e}^+_F + {\bf e}^-_F)
\ \ \ \ \& \ \ \ \ \frac{ d{\bf e}^\pm_F}{ds}  = -  \kappa {\bf t} \mp i \tau
{\bf e}^{\pm}_F
\]
Here $\kappa(s)$ is the extrinsic curvature and $\tau(s)$ is
the torsion. They specify the extrinsic geometry of the string:
Once $\kappa(s)$ and $\tau(s)$ are known the shape of the string can be 
constructed by solving the Frenet equations. The solution is 
defined uniquely in $\mathbb R^3$ up to rigid Galilean motions. 

The concept of gauge invariance emerges
from the following simple observation \cite{oma}: The vectors $\bf n$ 
and $\bf b$ span the normal plane of the string. But any physical 
property of the string must be independent of the choice of basis 
on the normal plane, and instead of ${\bf e}^\pm_F$ we 
could introduce another frame which is related to the Frenet 
frame by a rotation with an angle $\theta(s)$ on the normal plane,
\[
{\bf e}^+_F \to e^{ i \theta} {\bf e}^+_F \equiv
{\bf e}^+_\theta .
\]
When we substitute this the Frenet equation we conclude that the
U(1) rotation redefines 
\be
\begin{matrix}
\kappa \to e^{i\theta} \kappa \equiv \kappa_\theta  \\ 
\tau \to \tau + \partial_s \theta \equiv \tau_\theta 
\end{matrix}
\label{kt}
\ee
In the relations (\ref{kt}) we identify the gauge transformation 
structure of two dimensional Abelian Higgs multiplet $(\phi, A_i)$. 
The frame rotation corresponds to
a static U(1) gauge transformation, $\kappa_\theta$ corresponds to the 
complex scalar field $\phi \sim \kappa_\theta$, and $\tau_\theta$ corresponds
to the spatial component $A_1 \sim \tau_\theta$ of the U(1) gauge field. 

Since the physical properties of the string are independent of the choice
of a local frame, they must remain invariant under the U(1) 
transformation (\ref{kt}).
In particular, any effective Landau-Ginzburg
energy functional that describes a homopolymer and involves
the multiplet $(\kappa_\theta, \tau_\theta) \sim (\phi, A_1)$ must 
be U(1) gauge invariant.

The following variant of the Abelian Higgs model Hamiltonian \cite{landau}
is {\it the}  natural choice for a gauge invariant (internal)
Landau-Ginzburg energy functional, 
\[
F \ = \ \int\limits_0^L \! \! ds \, \left\{ 
\, |(\partial_s -  i A_1) \phi|^2 +
c \, ( |\phi|^2 - \mu^2)^2 \, \right\}  
\]
\be
 + \ d  \!
\int\limits_0^L \!\! ds \, A_1 .
\label{ahm}
\ee
Here the first term is the conventional energy functional of the Abelian Higgs
model including a gauge invariant kinetic and potential terms. For simplicity
we choose the Higgs potential so that it has the canonical quartic 
functional form. When $\mu \not= 0$ and real valued 
we have a spontaneous symmetry breaking with the ensuing
Higgs effect, and the ground state of the string acquires
a nonvanishing local curvature. The last term is
the one-dimensional version of the Chern-Simons functional \cite{chern}. We 
shall find that its presence provides a very simple explanation of 
homochirality, with a negative (positive) parameter $\sigma$ giving rise 
to right-handed (left-handed) chirality.  

We determine the thermodynamical properties of (\ref{ahm}) from the canonical partition
function, defined in the usual manner by integrating over the fields $\phi$ and $A_1$ 
\[
Z \ = \ \int [d\phi] [dA_1] \exp \{ - \int\limits_{0}^\beta \! d\tau \,  F(\phi, A_1) \} 
\]
We take the measure to be the canonical measure in the $(\phi, A_1)$ space (with
appropriate gauge fixing). Alternatively
we could also introduce the canonical (Polyakov) measure in the coordinate space $\bf r$, and
the two measures differ by a Jacobian factor $\mathcal J [ \phi, A]$ that appears as 
a correction to the free energy (\ref{ahm}),
\[
F \ \to \ F + \int\limits_{0}^L \! ds \, \ln [ \mathcal J ]
\]
The Jacobian is in general a non-local functional of $\phi(s)$ and $A_1(s)$ and we do not have
its general form at our disposal. But we can expand
it in power of the derivatives of these variables: Since the Jacobian is gauge invariant, by general
arguments of gauge invariance to lowest nontrivial order in $\phi$ and $A_1$ the result must have the same functional 
as the terms that we have already included in (\ref{ahm}). Consequently at the the present level
of approximation  we strongly suspect that the the {\it only} effect
of the Jacobian  would be to renormalize the parameters that already appear in (\ref{ahm}).  It
would be very interesting to study this issue in more detail.  

\section{discretization}

We now proceed to the discrete lattice version of (\ref{ahm}) that we use in our actual
computations. We first eliminate the explicit gauge dependence by implementing
the invertible change of variables
\[
\phi \leftrightarrow \rho e^{i\psi}
\]
\[
J \leftrightarrow \frac{1}{2i|\phi|^2} \{ 2iA_1 |\phi|^2 - \phi^\star \partial_s \phi + c.c. \}
\]
where the gauge invariant variable $J$ is called the supercurrent in the context of superconductivity.
The Jacobian for this change of variables is $\rho$.
With the identifications $\rho \to \kappa$ and $J \to \tau$ we then arrive at the
following discrete  version of the free energy (\ref{ahm})  to describe
the properties of  general chiral polymers,
\[
F \ = \ 
\sum_{i,j=1}^{N} a_{ij} \left\{ 1-\cos[\omega_{ij}(\kappa_i -
\kappa_{j})] \right\} \]
\be
+ \sum_{i=1}^{N}  \left\{ b_i \kappa_i^2 \tau_i^2 + c_i \cdot
(\kappa_i^2 - \mu_i^2)^2 \right\} \ + \ \sum_{i=i}^{N} d_i  \tau_i 
\label{dahm}
\ee
The $i,j=1,...,N$ label the monomers, and $a_{ij}, \omega_{ij}$ and $b_{ij}$ are parameters
that we have normalized to unity in (\ref{ahm}) but now included for completeness;  Note that
as we write it, there is  a superfluous overall scale in (\ref{dahm}).  
The first term describes long-distance correlations, it is the discrete analog of
the derivative term of the Higgs field in the continuum limit. 
We have introduced the cosine function to tame excessive fluctuations 
in $\kappa_i$. The middle term describes the interaction between
$\kappa_i$ and $\tau_i$, and the symmetry breaking 
self-interaction of $\kappa_i$. Finally, the last term 
is a discretized one-dimensional version of the Chern-Simons 
functional \cite{chern} that  is the origin of  homochirality.

In addition, one could also add the Jacobian that emerges
from the supercurrent change of variables. We have tested our model 
with this Jacobian included and we have found that it has no essential 
qualitative consequences, thus it will be excluded from the present analysis.

We relate the dynamical variables $(\kappa_i, \tau_i)$ 
in (\ref{dahm}) to the polymer 
geometry as follows: The modulus of the Higgs field $\kappa_i$ we identify with
the signed Frenet curvature of the backbone at the site $i$, and 
$\tau_i$ is the corresponding Frenet torsion. 
Once the numerical values of $\kappa_i$ and $\tau_i$ 
are known, the geometric shape of the polymer in the three dimensional
space $\mathbb R^3$ is obtained by integrating a discretized 
version of the Frenet equations. This integration
also introduces parameters $\Delta_i$, the average finite 
distance between the monomers.

For a general polymer the quantities ($a_{ij},\omega_{ij}, b_i, c_i,$ $\mu_i,d_i$) are {\it a priori} free site-dependent
parameters,  and different values of these parameters can be used to describe 
different kind of monomer structures.  Here we shall be interested in the limiting case
of {\it homopolymers}, where  we restrict ourselves to only the nearest neighbor interactions
with
\be
a_{ij} = \left\{  \begin{matrix}
\hskip -1.2cm 
a \cdot ( \delta_{i, i+1} + \delta_{i, i-1} ) \ \ \ \ (i=2, ... ,N-1)
\\
\hskip 0.2cm a \ \ \ \ \ \ \ \ (i=1,j=2) \ \ \& \ \ (i=N-1,j=N) \end{matrix}
\right.
\label{a}
\ee
and we also select {\it all} the remaining parameters to be
{\it independent } of the site index $i$. 

\section{numerical simulations}

We have employed  (\ref{dahm}) to study 
polymer collapse at low temperatures using
Monte Carlo free energy minimization, 
in the limiting case  of a homopolymer where all the parameters
are site-independent; see  (\ref{a}).  
At each iteration step of the numerical energy minimization procedure 
we first generate a new set of
values for the curvature and torsion $(\kappa_i, \tau_i)$ using the 
Metropolis algorithm \cite{metro} with a finite Metropolis 
temperature-like parameter $T_M$.
We then construct a new polymer configuration  by solving the 
discrete Frenet equations with a fixed and uniform distance between monomers
$\Delta$,
\be
|{\bf r} (s_i) - {\bf r}(s_{i-1}) | = \Delta \ \ \ \ \ \ i=2,...,N .
\label{cond1}
\ee
Finally, before accepting the new configuration
we exclude steric clashes by demanding that the distance 
between any two monomers  in the new configuration satisfies the bound
\be
|{\bf r} (s_i) - {\bf r}(s_j) | \geq z  \ \ \ \ \ {\rm   for} \ \ \ 
|i-j| \geq 2 .
\label{cond2}
\ee

Our simulations start from an initial configuration with $\kappa_i = 
\tau_i = 0$. This corresponds to a straight, untwisted polymer. 
Since the initial Metropolis step is determined randomly, essentially
by a thermal fluctuation, our starting point has a 
large conformational entropy. Consequently we expect 
that statistically our final conformations cover a substantial
portion of the landscape of  collapsed polymers. 

Depending on application
we can derive restrictions on the parameters, 
for example by comparing the results of our simulations to the properties of actual
polymers.  In the biologically interesting case where the 
model is used to describe statistical properties of folded structures  in the Protein
Data Bank (PDB) {\cite{pdb}, we would impose the constraint that in 
a full $2\pi$ $\alpha$-helix turn 
there are on average about 3.6 monomers (central $\alpha$ carbons).

We have made extensive numerical simulations using configurations where
the number $N$ of monomers lies in the range $75 \leq 
N \leq 1,000$. For these configurations we typically arrive at a stable 
collapsed state after around 1,000,000 steps. The folding process takes 
no more than a few tens of seconds in a MacPro desktop computer, even 
for the large values of $N$. But in order to ensure the stability 
of our final configurations we have extended our simulations to 22,000,000 
steps. Besides thermal fluctuations, we observe no essential change in 
the collapsed structures after the initial 1,000,000 steps which confirms that
we have reached a native state.   

\section{comparison with proteins}

We have compared the predictions of the homogeneous limit of the model (\ref{dahm}) to the
statistical properties of protein structures that have been deposited in the Protein Data Bank. 
The protein backbones all have
an identical homogeneous structure.  But we recognize that the detailed fold of a {\it given} protein 
is presumed to be  strongly influenced by  the specifics of the interactions that involve its unique amino acid
sequence. These include hydrophobic, 
hydrophilic, long-range Coulomb, van der Waals,  saturating hydrogen bonds  {\it etc.} interactions. Consequently
a given protein should not be approximated by a homopolymer model. However, one can argue that
when one asks questions that relate to the common statistical properties of all proteins that are stored in
the PDB  one can expect that the inhomogeneities that are due to the 
different amino acid structures  become less relevant and statistically, in average,  these proteins behave very much
like a homopolymer.
We find it  interesting to try and see whether this kind of argumentation is indeed correct.

In Figure 1 we have placed all 
single-stranded proteins that can be presently harvested
from the Protein Data Bank, with 
the number $N$ of central carbons in the range of $75 \leq N \leq 1,000$. 
Using a least square linear fit to the data we find for the compactness index 
the value $\nu_{PDB} \approx  0.378 \pm 0.0017$, which is in line with the results 
previously reported in the literature \cite{nu2}, \cite{greg}. 
In Figure 1 we also show how the compactness index $\nu$ 
in our model depends on  $N$
when $75 \leq N \leq 1,000$, using a statistical sample of 80 runs
for each value of $N$. When we apply a least square 
linear fit to our results we find for the compactness index
the estimate $\nu \approx  0.379 \pm 0.0081$, a somewhat surprisingly 
excellent agreement 
with the value obtained from the
Protein Data Bank. Since $\nu$ is a universal quantity, this agreement
implies  that in the sense of the compactness index our model  resides in the 
same universality class with proteins in PDB.
\begin{figure}[h]
        \centering
\includegraphics[width=0.4\textwidth]{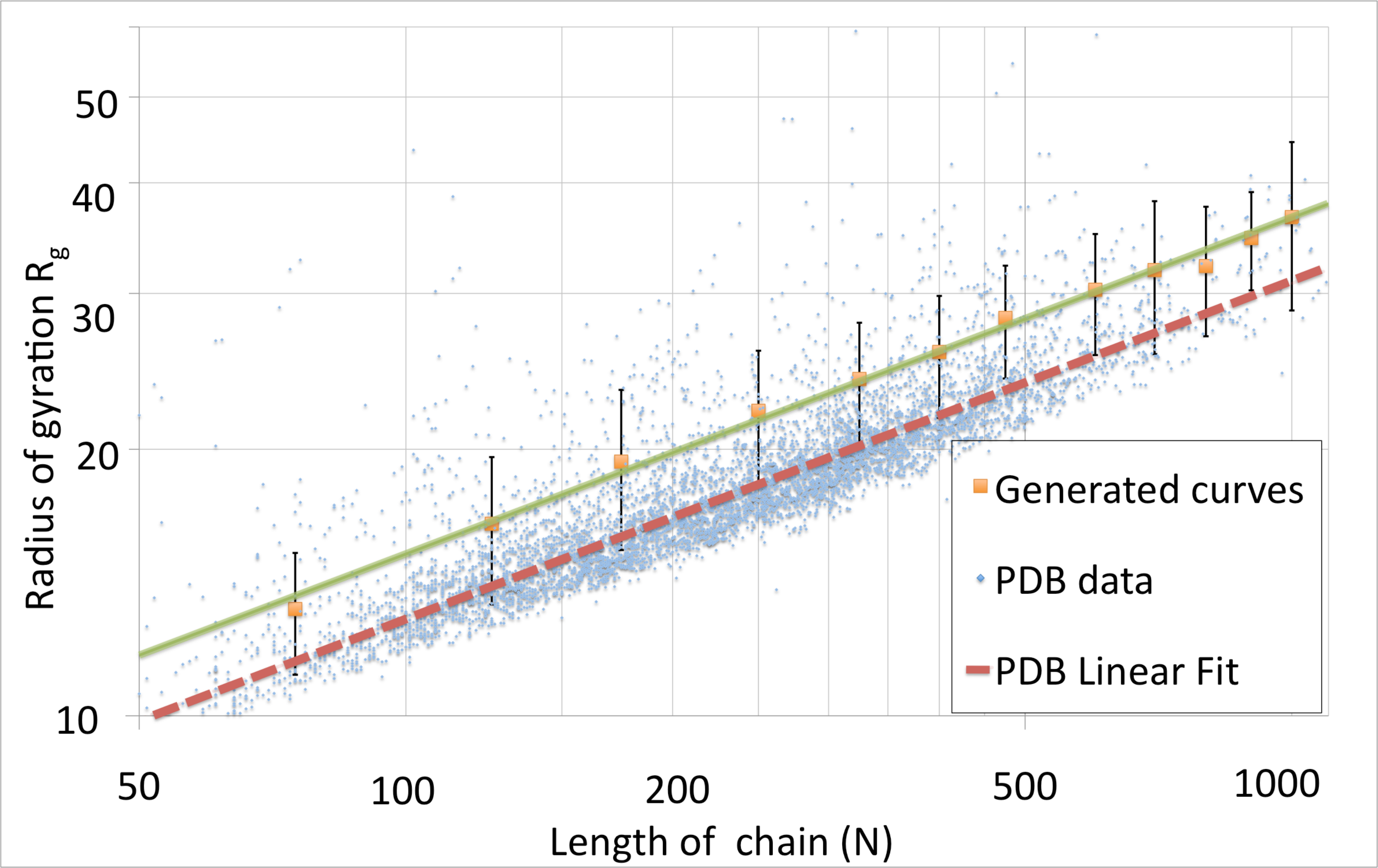}
\caption{\it (Color online) Least square linear fit to the 
             compactness index $\nu$ computed in our model
       ($\nu \approx 0.379 \pm 0.0081$) compared 
        with that describing all single 
       strand proteins currently deposited at the Protein Data Bank 
       ($\nu_{PDB} \approx  0.378 \pm 0.0017$). The error-bars 
       describe standard deviation from the average, and it can be viewed
       as a measure of 
       conformational entropy in our configurations. 
         }
        \label{Figure 1:}
\end{figure}
\noindent

From the data in Figure 1 we find that our model predicts
for the form factor $L$ in (\ref{nu})   
the numerical value $L \approx 2.656 \pm 0.049 $ (\.A). 
This compares well with the average value $L_{PDB} 
\approx 2.254 \pm 0.021 $ (\.A) that we obtain using a least 
square fit to the Protein Data Bank data displayed in Figure 1.

We note that unlike the compactness index $\nu$, the value of $L$ is not a universal
quantity and its value can be influenced by varying the parameters.
The explicit parameter values that we have used in our simulations are
$a=4$, $\omega = 4.25$, $b=0.0005488$, $c=0.5$, $\mu = 24.7$,
$d=-20$, and  $\Delta = 3.8$, $z=3.7$ and  we have selected these parameter 
values by trial-and-error to
produce a value for $L$ that is as close as possible to the experimental value.
We have verified that by changing the parameter values
$\nu$ remains  in the vicinity of $\nu \approx 0.38$ while $L$ can change substantially.
 
We observe from Figure 1 that the standard deviation 
displayed by our final conformations 
are comparable in size to the actual spreading 
of PDB proteins around their experimentally 
determined average values. Since this standard deviation is a measure of
conformational entropy, we conclude that  at each value of $N$ 
our initial configuration appears to have enough conformational entropy
to cover the entire landscape of native state protein
folds in PDB. 

We have verified that in our model the value of $\nu$ is temperature 
independent for a wide range of temperatures: The value of  $\nu$ is
{\it insensitive} to an increase in the Metropolis temperature 
$T_M$ until $T_M$ reaches a critical value $T_1$. At this critical temperature there is an 
onset of a transition towards the $\Theta$-point, and
at the $\Theta$-point we estimate $\nu \approx 0.48 - 0.49$ in line 
with the expected value $\nu \sim  1/2$ that characterizes
the universality class of a random coil. In the limit of high 
temperatures we find $\nu \approx 0.65 $ which is slightly above but in line with
the Flory value $\nu = 3/5$ for a self-avoiding random walk.

We have also studied the effect of the various operators in  
(\ref{dahm}) in determining the universality class:

We find indications that the value $\nu \approx 0.38 $ 
is driven  by the presence of the chirality 
breaking Chern-Simons term: When we entirely remove the Chern-Simons term 
by setting $d_i=0$ while keeping all other parameters intact in (\ref{dahm}), we find that
the compactness index increases to $\nu \approx 0.488 \dots$ which is very close to 
the $\Theta$-point value $\nu \sim 1/2$. This suggests that according to our model
there is some  relation between chirality and the  transition to the collapsed phase in the
case of homopolymers, that deserves to be investigated in more detail.

When we in addition remove the direct coupling between torsion
and curvature by setting $b_i = d_i = 0$ the compactness index remains
near its $\Theta$-point value $\nu \approx 0.488 \dots$.

Finally, we have observed that when we remove the entire symmetry breaking potential
by setting $c_i = 0$ in (\ref{dahm}) we find that $\nu$ approaches
the value $\nu \approx 0.737 \dots$. This proposes that we 
may have a novel universality class in the low temperature
phase. Notice that in this case the local minima of
the potential energy are absent and the discrete symmetry 
$\phi \to -\phi$ becomes restored at the classical level. 

At a very high Metropolis temperature $T_M$ and when we set $b_i = d_i = 0$
we find that the compactness index, as expected, approaches the 
Flory value 3/5; we now get $\nu \approx 0.61 \dots$. 

\vskip 0.3cm
In Figure 2 we show using an example with $N=300$, 
how the compactness index $\nu$ evolves as a function
of the number of iterations ("time"), during the first 1,000,000 steps. 
In this figure we also describe how the free energy (\ref{dahm})
develops as a function of the iteration steps.  
\begin{figure}[h]
        \centering
                \includegraphics[width=0.40\textwidth]{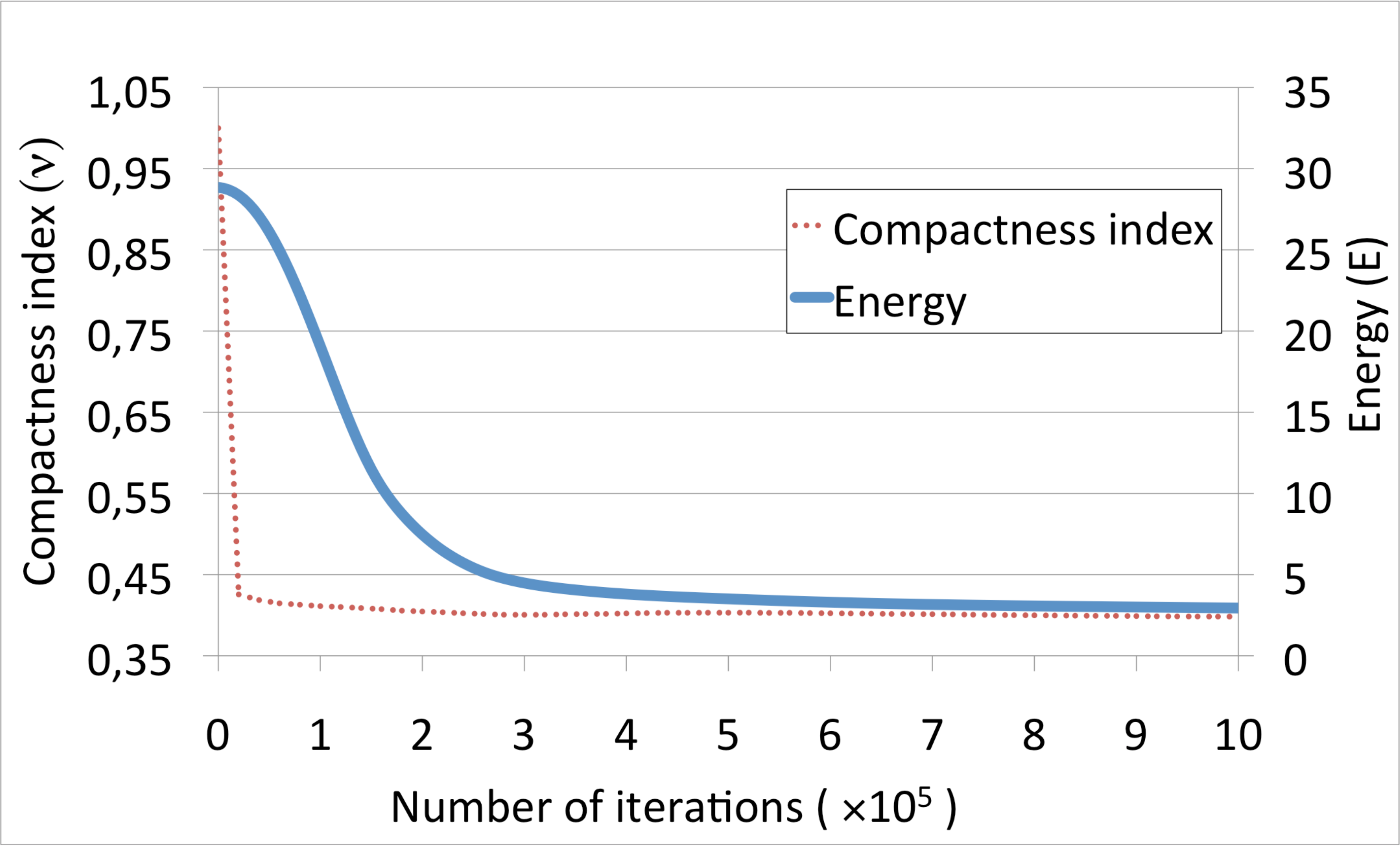}
\vskip 0.2cm
        \caption{{ \it (Color online) The dotted line shows how the 
compactness index $\nu$ typically evolves as a function 
        of the number of interation steps (time) when computed as an average
        over statistical samples and up to 1,000,000 steps.  
The continuous line shows similarly 
        how the average energy typically evolves as a 
function of the number of interation steps.
                }}
       \label{Figure 2:}
\end{figure}
We find that while $\nu$ generically approaches its asymptotic 
value $\nu \approx 0.38 $ very rapidly, after only a 
few thousand iterations, the 
process of energy minimization 
typically takes about two orders of magnitude longer. 
The asymptotic behaviour of the
curves confirms that the final state is highly stable. 
The stability is further validated 
by a comparison with Figure 1 where we report on results 
after the iteration process 
has been continued by 21,000,000 additional 
steps: For $T_M < T_\Theta $  we find
no essential change in the final conformations after $N \sim 1,000,000$
steps, beyond thermal fluctuations.

While we realize that our Monte Carlo simulation is not designed to be
a reliable method for describing the out-of-equilibrium dynamical 
time evolution of polymer folding,  we still find it curious that according to 
Figure 2 the process of collapse as described by our model is very 
much like the expected folding process of biological proteins: 
The initial denatured state first rapidly collapses into a molten 
globule, with a large decrease in conformational entropy but only a
very small change in the internal energy. After the initial collapse to 
the molten globule with the ensuing 
formation of secondary structures such as $\alpha$-helices and 
$\beta$-sheets, the  process continues with a relatively 
slow conformational re-arrangement towards a locally stable 
conformation. The 
final state has a substantially lower energy than the 
corresponding molten globule state.

Finally, we have also compared the geometrical shape of the collapsed configurations as computed
in our model to that of folded proteins using the hierarchical
classification scheme CATH \cite{cath}. We find that the geometry of
the configurations computed using our model 
are in a very good correspondence with this classification scheme, and
they look very much like actual folded proteins.
In particular, our model appears to produce all the major 
secondary structures of proteins in PDB;
See figure 4 for typical examples.
\begin{figure}[h]
        \centering
                \includegraphics[width=0.45\textwidth]{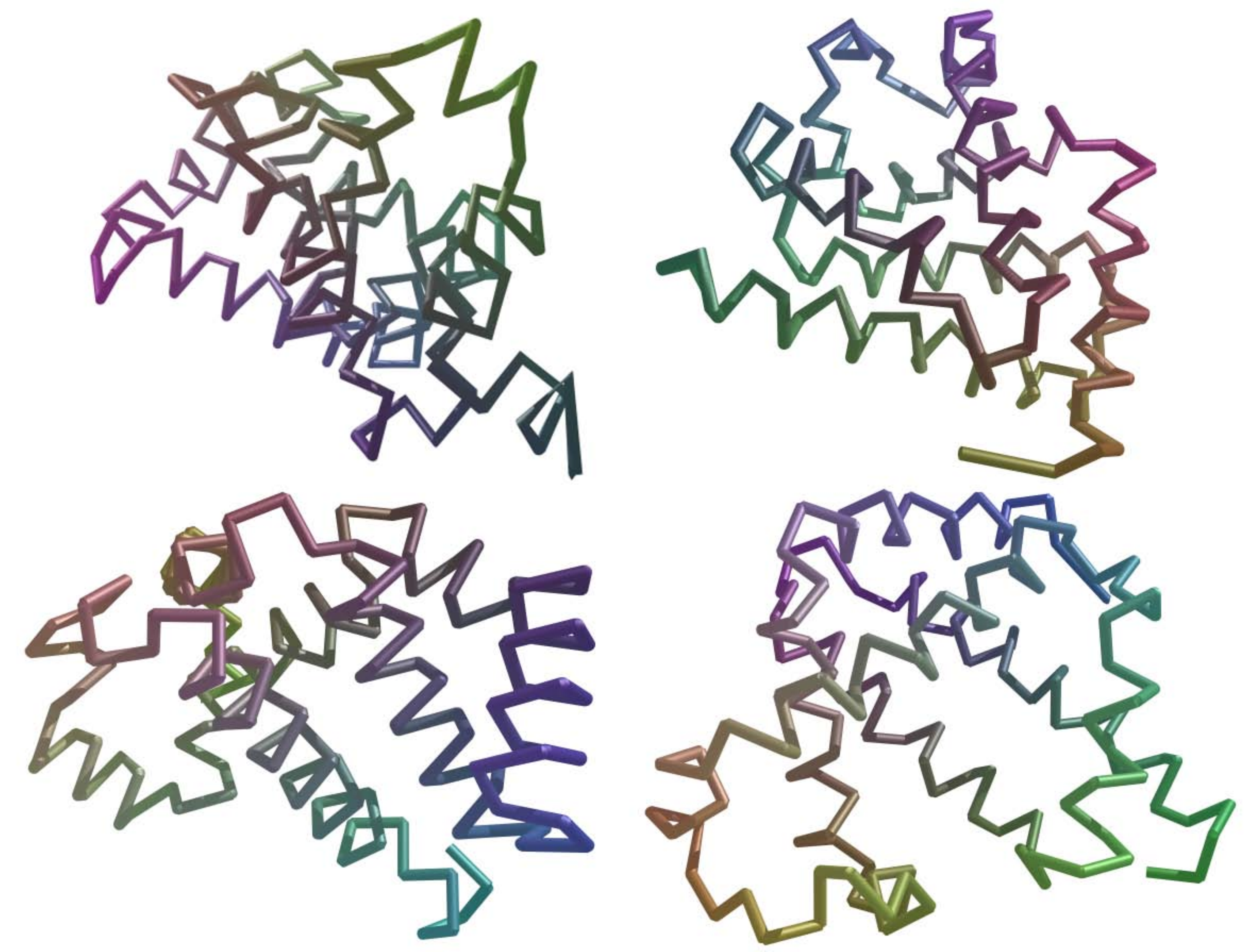}
        \caption{\it (Color online) On the left two {\it a priori} generic examples of folded configurations computed in our model
        with $N=153$.  Motifs such as helicity-loop-helicity are clearly visible.
        On the right, for comparison, are pictures of myoglobin 1mbn (above) and 1m6c (below)
        backbones, constructed using data taken from Protein Data Bank.}
             \label{Figure 3}
\end{figure}

\section{summary}

In summary, we have developed an effective 
field theory model that describes the collapse of  a  homopolymer
in its low temperature phase. We have compared various properties
of our model with the statistical, average properties of folded
proteins that are stored in the Protein Data Bank. We have found that our model
reproduces the statistical properties of the PDB proteins with surprisingly
good accuracy. For example, it computes accurately the compactness 
index $\nu$ of native state proteins and correctly describes the phenomenology of
protein collapse. Furthermore,  since the folded  states 
obtained in our model are also in line with the CATH classification scheme, 
it appears that our model has promise for  a tool to 
analyze the statistical properties of folded proteins. 

\vskip 0.3cm

Our research is supported by grants from the Swedish Research Council (VR).
The work by A.J.N is also supported by the Project Grant
ANR NT05-142856. A.J.N. thanks H. Orland for discussions and
advice. We all thank M. Chernodub for discussions, 
and N. Johansson and J. Minahan for comments.
A.J.N. also thanks T. Gregory Dewey for communications. 
A.J.N. thanks the Aspen Center for 
Physics for hospitality during this work.

\end{document}